\pdfoutput=1
\documentclass[12pt,a4paper]{article}

\usepackage{ifthen} 
\usepackage{rotating}
\usepackage{dashrule}
\usepackage{array} 
\usepackage{dcolumn}
\usepackage{comment}
\usepackage{xcolor}

\usepackage{graphpap} 
\usepackage{makecell} 
\usepackage{rotating} 
\usepackage{tikz}
\usepackage{xcolor}
\usepackage{longtable} 

\usepackage{lscape}
\usepackage{multirow} 
\usepackage{mathrsfs}
\usepackage{mathtools} 
\usetikzlibrary{patterns}
\usepackage{nicefrac} 
\usepackage{transparent}
\usepackage{afterpage}

\newboolean{pdflatex}
\setboolean{pdflatex}{true} 

\newboolean{articletitles}
\setboolean{articletitles}{true} 

\newboolean{uprightparticles}
\setboolean{uprightparticles}{True} 

\def\paperauthors{LHCb collaboration} 
\def\paperasciititle{Search for the Bc+ -> X(3872)pi decay} 
\def\papertitle{Search for the \BcToXpi decay} 

\def\paperkeywords{{High Energy Physics}, {LHCb}} 
\def\papercopyright{\the\year\ CERN for the benefit of the LHCb collaboration} 
\def\paperlicence{CC BY 4.0 licence}
\def\paperlicenceurl{https://creativecommons.org/licenses/by/4.0/}

\usepackage{wasysym}

\DeclareMathOperator*{\bigplus}{\scalerel*{+}{\sum}}
\usepackage{scalerel}

\makeatletter
\g@addto@macro\bfseries{\boldmath}
\makeatother

\newif\ifEnableSectionTOCLinks
\EnableSectionTOCLinksfalse 

\usepackage[top=1in, bottom=1.25in, left=1in, right=1in]{geometry}

\usepackage{microtype}
\usepackage{lineno}  
\usepackage{xspace} 
\usepackage{caption} 

\usepackage{graphicx}  
\usepackage{color}
\usepackage{colortbl}
\graphicspath{{./figs/}} 

\usepackage{amsmath} 
\usepackage{amssymb}
\usepackage{amsfonts}
\usepackage{upgreek} 

\newcommand*\patchAmsMathEnvironmentForLineno[1]{%
\expandafter\let\csname old#1\expandafter\endcsname\csname #1\endcsname
\expandafter\let\csname oldend#1\expandafter\endcsname\csname
end#1\endcsname
 \renewenvironment{#1}%
   {\linenomath\csname old#1\endcsname}%
   {\csname oldend#1\endcsname\endlinenomath}%
}
\newcommand*\patchBothAmsMathEnvironmentsForLineno[1]{%
  \patchAmsMathEnvironmentForLineno{#1}%
  \patchAmsMathEnvironmentForLineno{#1*}%
}
\AtBeginDocument{%
\patchBothAmsMathEnvironmentsForLineno{equation}%
\patchBothAmsMathEnvironmentsForLineno{align}%
\patchBothAmsMathEnvironmentsForLineno{flalign}%
\patchBothAmsMathEnvironmentsForLineno{alignat}%
\patchBothAmsMathEnvironmentsForLineno{gather}%
\patchBothAmsMathEnvironmentsForLineno{multline}%
\patchBothAmsMathEnvironmentsForLineno{eqnarray}%
}

\usepackage[pdftex]{hyperref}
\usepackage{hyperxmp}
\hypersetup{pdfauthor={\paperauthors},
            pdftitle={\paperasciititle},
            pdfkeywords={\paperkeywords},
            pdfcopyright={Copyright (C) \papercopyright},
            pdflicenseurl={\paperlicenceurl}}
\usepackage[colorinlistoftodos,textsize=scriptsize]{todonotes}

\usepackage[bottom,flushmargin,hang,multiple]{footmisc}

\usepackage[all]{hypcap} 

\setboolean{uprightparticles}{true} 
\usepackage{xspace} 
\usepackage{upgreek}

\def\lhcb   {\mbox{LHCb}\xspace}
\def\atlas  {\mbox{ATLAS}\xspace}
\def\cms    {\mbox{CMS}\xspace}

\def\babar  {\mbox{BaBar}\xspace}
\def\belle  {\mbox{Belle}\xspace}

\def\besiii {\mbox{BESIII}\xspace}

\def\cdf    {\mbox{CDF}\xspace}
\def\dzero  {\mbox{D0}\xspace}

\def\MagUp {\mbox{\em Mag\kern -0.05em Up}\xspace}

\ifthenelse{\boolean{uprightparticles}}%
{

 \def\Pmu         {\ensuremath{\upmu}\xspace}

 \def\Ppi         {\ensuremath{\uppi}\xspace}                 
                  
 \def\Prho        {\ensuremath{\uprho}\xspace}

 \def\Pphi        {\ensuremath{\upphi}\xspace}                 
                  
 \def\Pchi        {\ensuremath{\upchi}\xspace}                 
 \def\Ppsi        {\ensuremath{\uppsi}\xspace}

 \def\PDelta      {\ensuremath{\Delta}\xspace}                 
 \def\PXi         {\ensuremath{\Xi}\xspace}                 
 \def\PLambda     {\ensuremath{\Lambda}\xspace}                 
 \def\PSigma      {\ensuremath{\Sigma}\xspace}                 
 \def\POmega      {\ensuremath{\Omega}\xspace}                 
 \def\PUpsilon    {\ensuremath{\Upsilon}\xspace}

 \def\PB      {\ensuremath{\mathrm{B}}\xspace}                 
                  
 \def\PD      {\ensuremath{\mathrm{D}}\xspace}

 \def\PH      {\ensuremath{\mathrm{H}}\xspace}                 
                  
 \def\PJ      {\ensuremath{\mathrm{J}}\xspace}                 
 \def\PK      {\ensuremath{\mathrm{K}}\xspace}

 \def\PP      {\ensuremath{\mathrm{P}}\xspace}

 \def\PS      {\ensuremath{\mathrm{S}}\xspace}

 \def\PW      {\ensuremath{\mathrm{W}}\xspace}                 
 \def\PX      {\ensuremath{\mathrm{X}}\xspace}

 \def\Pb      {\ensuremath{\mathrm{b}}\xspace}                 
 \def\Pc      {\ensuremath{\mathrm{c}}\xspace}                 
                  
 \def\Pe      {\ensuremath{\mathrm{e}}\xspace}

 \def\Pi      {\ensuremath{\mathrm{i}}\xspace}

 \def\Pp      {\ensuremath{\mathrm{p}}\xspace}                 
 \def\Pq      {\ensuremath{\mathrm{q}}\xspace}                 
                  
 \def\Ps      {\ensuremath{\mathrm{s}}\xspace}

 \def\thebaroffset{0.0em}
}
{

 \def\Pmu         {\ensuremath{\mu}\xspace}

 \def\Ppi         {\ensuremath{\pi}\xspace}                 
                  
 \def\Prho        {\ensuremath{\rho}\xspace}

 \def\Pphi        {\ensuremath{\phi}\xspace}                 
                  
 \def\Pchi        {\ensuremath{\chi}\xspace}                 
 \def\Ppsi        {\ensuremath{\psi}\xspace}                 
                  
 \mathchardef\PDelta="7101
 \mathchardef\PXi="7104
 \mathchardef\PLambda="7103
 \mathchardef\PSigma="7106
 \mathchardef\POmega="710A
 \mathchardef\PUpsilon="7107
                  
 \def\PB      {\ensuremath{B}\xspace}                 
                  
 \def\PD      {\ensuremath{D}\xspace}

 \def\PH      {\ensuremath{H}\xspace}                 
                  
 \def\PJ      {\ensuremath{J}\xspace}                 
 \def\PK      {\ensuremath{K}\xspace}

 \def\PP      {\ensuremath{P}\xspace}

 \def\PS      {\ensuremath{S}\xspace}

 \def\PW      {\ensuremath{W}\xspace}                 
 \def\PX      {\ensuremath{X}\xspace}

 \def\Pb      {\ensuremath{b}\xspace}                 
 \def\Pc      {\ensuremath{c}\xspace}                 
                  
 \def\Pe      {\ensuremath{e}\xspace}

 \def\Pi      {\ensuremath{i}\xspace}

 \def\Pp      {\ensuremath{p}\xspace}                 
 \def\Pq      {\ensuremath{q}\xspace}                 
                  
 \def\Ps      {\ensuremath{s}\xspace}

 \def\thebaroffset{0.18em}
}
\newcommand{\offsetoverline}[2][\thebaroffset]{\kern #1\overline{\kern -#1 #2}}%

\makeatletter
\ifcase \@ptsize \relax
  \newcommand{\miniscule}{\@setfontsize\miniscule{4}{5}}
\or
  \newcommand{\miniscule}{\@setfontsize\miniscule{5}{6}}
\or
  \newcommand{\miniscule}{\@setfontsize\miniscule{5}{6}}
\fi
\makeatother

\DeclareRobustCommand{\optbar}[1]{\shortstack{{\miniscule (\rule[.5ex]{1.25em}{.18mm})}
  \\ [-.7ex] $#1$}}

\def\epem       {{\ensuremath{\Pe^+\Pe^-}}\xspace}

\def\mumu       {{\ensuremath{\Pmu^+\Pmu^-}}\xspace}

\def\quark     {{\ensuremath{\Pq}}\xspace}
\def\quarkbar  {{\ensuremath{\overline \quark}}\xspace}

\def\squark    {{\ensuremath{\Ps}}\xspace}

\def\cquark    {{\ensuremath{\Pc}}\xspace}
\def\cquarkbar {{\ensuremath{\overline \cquark}}\xspace}
\def\ccbar     {{\ensuremath{\cquark\cquarkbar}}\xspace}
\def\bquark    {{\ensuremath{\Pb}}\xspace}
\def\bquarkbar {{\ensuremath{\overline \bquark}}\xspace}

\def\pion   {{\ensuremath{\Ppi}}\xspace}

\def\pip    {{\ensuremath{\pion^+}}\xspace}
\def\pim    {{\ensuremath{\pion^-}}\xspace}

\def\rhomeson {{\ensuremath{\Prho}}\xspace}
\def\rhoz     {{\ensuremath{\rhomeson^0}}\xspace}

\def\kaon    {{\ensuremath{\PK}}\xspace}

\def\KorKbar {\kern \thebaroffset\optbar{\kern -\thebaroffset \PK}{}\xspace}
\def\Kz      {{\ensuremath{\kaon^0}}\xspace}

\def\Kp      {{\ensuremath{\kaon^+}}\xspace}
\def\Km      {{\ensuremath{\kaon^-}}\xspace}

\def\KS      {{\ensuremath{\kaon^0_{\mathrm{S}}}}\xspace}

\def\Kstarz  {{\ensuremath{\kaon^{*0}}}\xspace}

\newcommand{\phiz}{\ensuremath{\Pphi}\xspace}

\def\D       {{\ensuremath{\PD}}\xspace}

\def\DorDbar {\kern \thebaroffset\optbar{\kern -\thebaroffset \PD}\xspace}
\def\Dz      {{\ensuremath{\D^0}}\xspace}

\def\Dstarp  {{\ensuremath{\D^{*+}}}\xspace}

\def\B       {{\ensuremath{\PB}}\xspace}

\def\BorBbar {\kern \thebaroffset\optbar{\kern -\thebaroffset \PB}\xspace}
\def\Bz      {{\ensuremath{\B^0}}\xspace}

\def\Bd      {{\ensuremath{\B^0}}\xspace}

\def\BdorBdbar {\kern \thebaroffset\optbar{\kern -\thebaroffset \Bd}\xspace}
\def\Bu      {{\ensuremath{\B^+}}\xspace}

\def\Bp      {{\ensuremath{\Bu}}\xspace}

\def\Bs      {{\ensuremath{\B^0_\squark}}\xspace}

\def\BsorBsbar {\kern \thebaroffset\optbar{\kern -\thebaroffset \Bs}\xspace}
\def\Bc      {{\ensuremath{\B_\cquark^+}}\xspace}

\def\x  {{\ensuremath{\PX{(3872)}}}\xspace}

\def\jpsi     {{\ensuremath{{\PJ\mskip -3mu/\mskip -2mu\Ppsi\mskip 2mu}}}\xspace}
\def\psitwos  {{\ensuremath{\Ppsi{\rm{(2S)}}}}\xspace}

\def\chicone  {{\ensuremath{\Pchi_{\cquark 1}}}\xspace}
\def\chictwo  {{\ensuremath{\Pchi_{\cquark 2}}}\xspace}

\def\Y#1S{\ensuremath{\PUpsilon{(#1S)}}\xspace}

\def\proton      {{\ensuremath{\Pp}}\xspace}

\def\Lz          {{\ensuremath{\PLambda}}\xspace}

\def\LorLbar     {\kern \thebaroffset\optbar{\kern -\thebaroffset \PLambda}\xspace}

\def\Lb           {{\ensuremath{\Lz^0_\bquark}}\xspace}

\def\chiconex{{\ensuremath{\chicone(3872)}}\xspace}

\def\BF         {{\ensuremath{\mathcal{B}}}\xspace}
\def\BR         {\BF}

\newcommand{\decay}[2]{\ensuremath{#1\!\to #2}\xspace} 

\def\to                 {\ensuremath{\rightarrow}\xspace}

\def\BcTojpsitripi    {\decay{\Bc}{\jpsi\pip\pip\pim}}

\def\BcToXpi    {\decay{\Bc}{\chiconex\pip}}
\def\BcToPpi    {\decay{\Bc}{\psitwos\pip}}

\def\BcTojpsipi       {\decay{\Bc}{\jpsi\pip}}
\def\BcToXccpi       {\decay{\Bc}{\PX_\ccbar\,\pip}}

\def\ChiconexTojpsipipi{\decay{\chiconex}{\jpsi\pip\pim}}

\def\JpsiPiPi{\jpsi\pip\pim}

\def\AT#1     {\ensuremath{A_{\mathrm{T}}^{#1}}\xspace}           

\def\C#1      {\ensuremath{\mathcal{C}_{#1}}\xspace}                       
\def\Cp#1     {\ensuremath{\mathcal{C}_{#1}^{'}}\xspace}                    
\def\Ceff#1   {\ensuremath{\mathcal{C}_{#1}^{\mathrm{(eff)}}}\xspace}        
\def\Cpeff#1  {\ensuremath{\mathcal{C}_{#1}^{'\mathrm{(eff)}}}\xspace}       
\def\Ope#1    {\ensuremath{\mathcal{O}_{#1}}\xspace}                       
\def\Opep#1   {\ensuremath{\mathcal{O}_{#1}^{'}}\xspace}                    

\newcommand{\nospaceunit}[1]{\ensuremath{\text{#1}}}       
\newcommand{\aunit}[1]{\ensuremath{\text{\,#1}}}       

\newcommand{\tev}{\aunit{Te\kern -0.1em V}\xspace}
\newcommand{\gev}{\aunit{Ge\kern -0.1em V}\xspace}
\newcommand{\mev}{\aunit{Me\kern -0.1em V}\xspace}
\newcommand{\kev}{\aunit{ke\kern -0.1em V}\xspace}
\newcommand{\ev}{\aunit{e\kern -0.1em V}\xspace}
\newcommand{\mevc}{\ensuremath{\aunit{Me\kern -0.1em V\!/}c}\xspace}
\newcommand{\gevc}{\ensuremath{\aunit{Ge\kern -0.1em V\!/}c}\xspace}
\newcommand{\mevcc}{\ensuremath{\aunit{Me\kern -0.1em V\!/}c^2}\xspace}
\newcommand{\kevcc}{\ensuremath{\aunit{ke\kern -0.1em V\!/}c^2}\xspace}
\newcommand{\gevcc}{\ensuremath{\aunit{Ge\kern -0.1em V\!/}c^2}\xspace}

\def\mum  {\ensuremath{\,\upmu\nospaceunit{m}}\xspace}

\def\fb   {\ensuremath{\aunit{fb}}\xspace}
\def\invfb   {\ensuremath{\fb^{-1}}\xspace}

\def\gsim{{~\raise.15em\hbox{$>$}\kern-.85em
          \lower.35em\hbox{$\sim$}~}\xspace}
\def\lsim{{~\raise.15em\hbox{$<$}\kern-.85em
          \lower.35em\hbox{$\sim$}~}\xspace}

\def\sqs   {\ensuremath{\protect\sqrt{s}}\xspace}

\def\pt         {\ensuremath{p_{\mathrm{T}}}\xspace}

\def\evtgen     {\mbox{\textsc{EvtGen}}\xspace}

\def\geant      {\mbox{\textsc{Geant4}}\xspace}

\def\photos     {\mbox{\textsc{Photos}}\xspace}

\def\pythia     {\mbox{\textsc{Pythia}}\xspace}

\def\tell1  {TELL1\xspace}
\def\ukl1   {UKL1\xspace}

\newcommand{\vs}{\mbox{\itshape vs.}\xspace}

\newcommand{\lhcborcid}[1]{\href{https://orcid.org/#1}{\hspace*{0.1em}\raisebox{-0.45ex}{\includegraphics[width=1em]{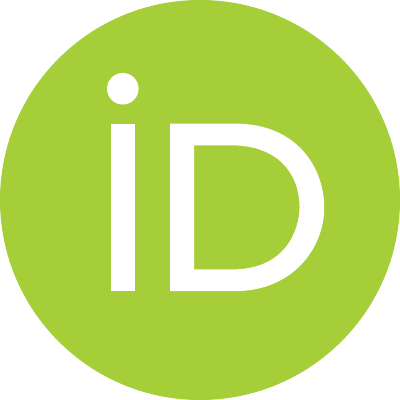}}}}

\hypersetup{
  colorlinks   = true, 
  urlcolor     = blue, 
  linkcolor    = blue, 
  citecolor    = red   
}

\ifEnableSectionTOCLinks
    \usepackage[explicit]{titlesec} 

    \let\oldcontentsline\contentsline
    \renewcommand

    \titleformat{\section}{\normalfont\Large\bf}{\hyperlink{tocsection.\thesection}{{\thesection} \parbox[t]{\dimexpr\textwidth-1pc}{#1}}}{1pc}{}

    \titleformat{\subsection}{\normalfont\bf}{\hyperlink{tocsubsection.\thesubsection}{{\thesubsection} \parbox[t]{\dimexpr\textwidth-1pc}{#1}}}{1pc}{}

    \titleformat{name=\section,numberless}[display]{}{}{0pt}{\normalfont\Huge\bfseries #1}
\fi

\usepackage{cite} 
\usepackage{mciteplus}

\usepackage{longtable} 

\newcolumntype{d}[1]{D{,}{\,\pm\,}{#1} }
\newcolumntype{f}[1]{D{,}{.}{#1} }

\begin{document}

\renewcommand{\thefootnote}{\fnsymbol{footnote}}
\setcounter{footnote}{1}


\begin{titlepage}
\pagenumbering{roman}

\vspace*{-1.5cm}
\centerline{\large EUROPEAN ORGANIZATION FOR NUCLEAR RESEARCH (CERN)}
\vspace*{1.5cm}
\noindent
\begin{tabular*}{\linewidth}{lc@{\extracolsep{\fill}}r@{\extracolsep{0pt}}}
\ifthenelse{\boolean{pdflatex}}
{\vspace*{-1.5cm}\mbox{\!\!\!\includegraphics[width=.14\textwidth]{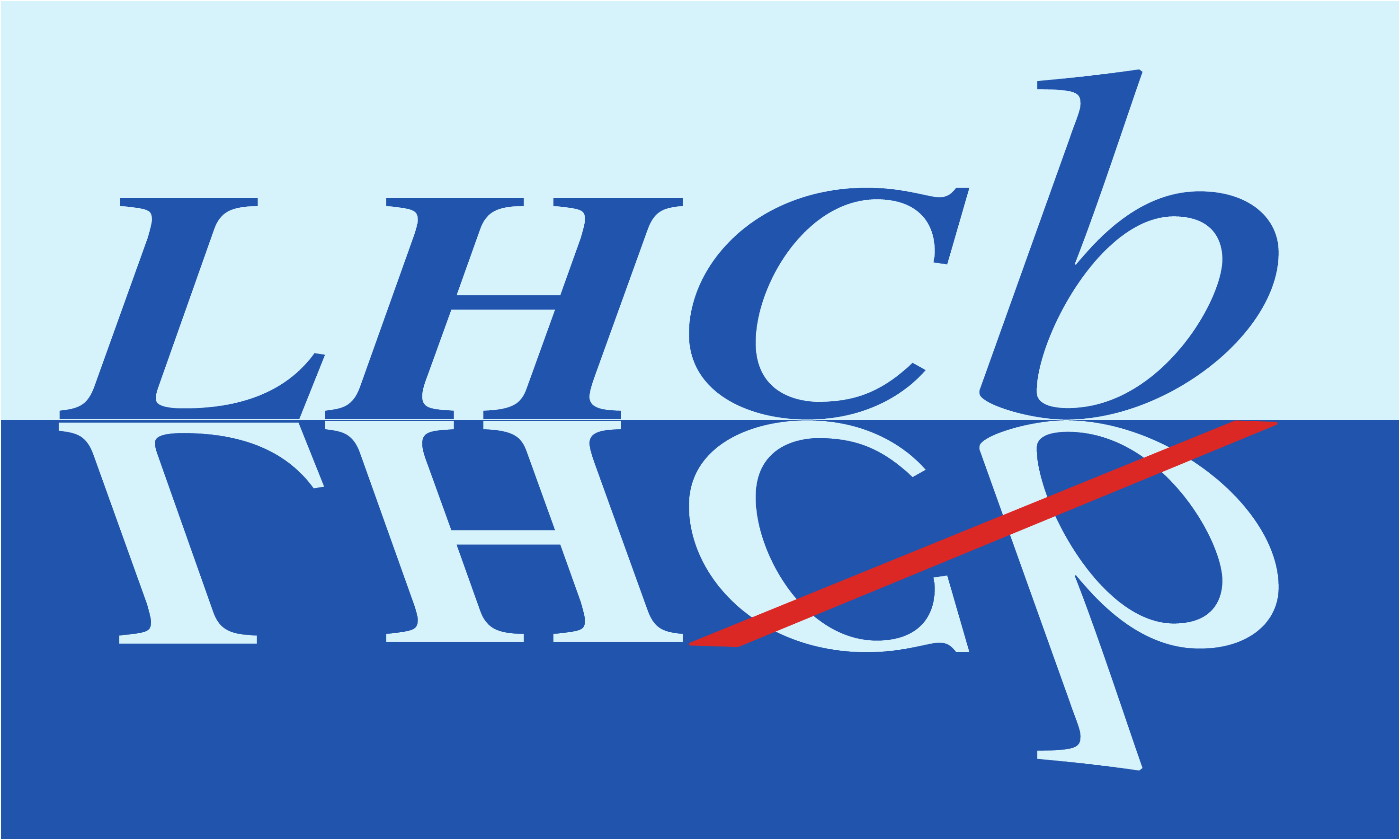}} & &}%
{\vspace*{-1.2cm}\mbox{\!\!\!\includegraphics[width=.12\textwidth]{figs/lhcb-logo.eps}} & &}%
\\
 & & CERN-EP-2025-042\\  
 & & LHCb-PAPER-2024-050 \\  
& & 24 March 2025 
\end{tabular*}

\vspace*{3.5cm}

{\normalfont\bfseries\boldmath\huge
\begin{center}
  \papertitle 
\end{center}
}

\vspace*{0.2cm}

\begin{center}
\paperauthors\footnote{Authors are listed at the end of this paper.}
\end{center}


\begin{abstract}
\noindent 
A~search for the~decay \mbox{$\decay{\Bc}{\chiconex\pip}$} 
is reported using proton\nobreakdash-proton collision data
collected with the~\lhcb detector 
between 2011 and 2018 at 
centre\nobreakdash-of\nobreakdash-mass energies of 
7, 8, and 13\tev, 
corresponding to an~integrated luminosity of~$9\invfb$.
No~significant signal 
is observed.
Using the~decay \mbox{$\decay{\Bc}{\psitwos\pip}$}
as a normalisation channel, 
an~upper limit for 
the~ratio of branching fractions  
$$
\mathcal{R}^{\chiconex}_{\psitwos}  = \dfrac{\BR_{\BcToXpi}}
      {\BR_{\BcToPpi}}
\times 
\dfrac{\BR_{\decay{\chiconex}{\jpsi\pip\pim}}}
      {\BR_{\decay{\psitwos}{\jpsi\pip\pim}}}
< 0.05\,(0.06)\,,
$$
is set at the~90\,(95)\% confidence level. 

\end{abstract}

\vspace{\fill}

\begin{center}
  Published in \href{https://doi.org/10.1007/jhep06(2025)013}{JHEP 06 (2025) 013}
\end{center}

\vspace*{0.5cm}

{\footnotesize 
\centerline{\copyright~\papercopyright. \href{\paperlicenceurl}{\paperlicence}.}}
\vspace*{2mm}

\end{titlepage}


\newpage
\setcounter{page}{2}
\mbox{~}
%

\cleardoublepage


\renewcommand{\thefootnote}{\arabic{footnote}}
\setcounter{footnote}{0}



\pagestyle{plain} 
\setcounter{page}{1}
\pagenumbering{arabic}


%

\section{Introduction}
\label{sec:Introduction}
The~\Bc mesons are 
of interest
because they contain two 
different heavy-flavour quarks, charm and beauty. 
The~ground state possesses 
a~diverse array of weak decay modes,
due to the~fact that either of the~heavy quarks
can decay while the~other acts as a~spectator quark.
Additionally, two valence quarks may annihilate via 
a~virtual \PW~boson.
Studies of \Bc~decay channels and measurements of their branching fractions 
contribute to improve the~understanding of
models describing strong interactions
and test various effective models. 
Experiments at the~Large Hadron Collider\,(LHC) have 
opened a~new era for~\Bc~meson investigations. 
The~large \bquark-quark production cross-section at 
the~LHC enables the~\atlas, \cms and \lhcb experiments 
to study the~production, decays and other properties 
of the~\Bc meson~\cite{LHCb-PAPER-2011-044, 
LHCb-PAPER-2012-054, 
LHCb-PAPER-2013-010, 
LHCb-PAPER-2013-021, 
LHCb-PAPER-2013-044, 
LHCb-PAPER-2013-047, 
LHCb-PAPER-2014-009, 
LHCb-PAPER-2014-039, 
LHCb-PAPER-2014-050, 
LHCb-PAPER-2014-060, 
CMS:2014oqy,
LHCb-PAPER-2015-024, 
ATLAS:2015jep,
LHCb-PAPER-2016-001, 
LHCb-PAPER-2016-022, 
LHCb-PAPER-2016-020, 
LHCb-PAPER-2016-055, 
LHCb-PAPER-2016-058, 
LHCb-PAPER-2017-035, 
LHCb-PAPER-2017-045, 
CMS:2017ygm,         
LHCb-PAPER-2019-033, 
LHCb-PAPER-2020-003, 
LHCb-PAPER-2021-023,
LHCb-PAPER-2021-034,
LHCb-PAPER-2022-025,
LHCb-PAPER-2023-037,
LHCb-PAPER-2023-039,
LHCb-PAPER-2023-046}. 

Over the past decade, numerous \Bc-meson decays involving 
conventional charmonium states (such as the~\jpsi, \psitwos or $\chictwo(1\PP)$~mesons) 
were observed~\cite{PDG2024}.
Decays of beauty hadrons have proven to be a~convenient and 
fruitful tool for the~search and study of 
charmonium\nobreakdash-like  states. These states exhibit properties that suggest 
the~presence of a~$\ccbar$~component in  their quark content, 
yet they cannot be associated with any conventional 
charmonium resonance~\cite{
Choi:2007wga,
Mizuk:2009da,
Chilikin:2013tch,
LHCb-PAPER-2014-014,
LHCb-PAPER-2015-038,
LHCb-PAPER-2016-015,
LHCb-PAPER-2016-018,
LHCb-PAPER-2016-019,
LHCb-PAPER-2018-034,
LHCb-PAPER-2018-043,
LHCb-PAPER-2020-035,
LHCb-PAPER-2020-044}.
The~first such state, the~$\chicone(3872),$\footnote{\x in the~original paper.} 
was observed by the~\belle collaboration
in the~$\jpsi\pip\pim$~mass spectrum from~\mbox{$\decay{\Bu}
{\jpsi\pip\pim\Kp}$}~decays~\cite{Belle:2003nnu}.
For over twenty years since its discovery,  
the~properties of this state have 
been intensively investigated in $\epem$ collisions by 
the~\babar~\cite{
BaBar:2004iez,
BaBar:2004oro,
BaBar:2006fjg,
BaBar:2007ixp,
BaBar:2007cmo,
Aubert:2008gu,
BaBar:2008flx,
BaBar:2010wfc},
\belle~\cite{
Belle:2003eeg,
Belle:2006olv,
Belle:2008fma,
Belle:2011vlx,
Belle:2011wdj,
Belle:2015qeg,
Belle:2022puc,
Belle:2023zxm},  
and \besiii~\cite{
BESIII:2013fnz,
BESIII:2015klj,
BESIII:2020nbj,
BESIII:2022bse,
BESIII:2023xta,
BESIII:2023hml}
collaborations.
Additionally, 
studies have been conducted in 
proton\nobreakdash-antiproton collisions 
by 
the~\cdf~\cite{CDF:2003cab,
CDF:2005cfq,
CDF:2006ocq,
CDF:2009nxk}
and \dzero~\cite{D0:2004zmu} collaborations,
as well as in 
proton\nobreakdash-proton\,($\proton\proton$)~collisions
by the~\atlas~\cite{ATLAS:2016kwu}, 
CMS~\cite{
CMS:2013fpt,
CMS:2020eiw}
and 
\lhcb~\cite{
LHCb-PAPER-2011-034,
LHCb-PAPER-2013-001,
LHCb-PAPER-2014-008,
LHCb-PAPER-2015-015,
LHCb-PAPER-2016-016,
LHCb-PAPER-2019-023,
LHCb-PAPER-2020-008,
LHCb-PAPER-2020-009,
LHcb-PAPER-2020-023,
LHCb-PAPER-2021-026,
LHCb-PAPER-2021-045,
LHCb-PAPER-2021-047,
LHCb-PAPER-2022-049,
LHCb-PAPER-2024-015,
LHCb-PAPER-2024-021}
collaborations.
The~LHCb collaboration 
has also explored 
proton\nobreakdash-nucleus collisions~\cite{LHCb-PAPER-2023-026},
and the~CMS collaboration 
has studied 
lead\nobreakdash-lead 
collisions~\cite{CMS:2021znk}.
However, this state has yet to be observed 
in exclusive $\Bc$\nobreakdash-meson decays.
The~study of $\Bc$\nobreakdash-meson decays 
into the~\chiconex~particle 
can be useful to clarify the~nature of this enigmatic state.
Within the~compact\nobreakdash-tetraquark interpretation of 
the~\chiconex~state\cite{Maiani:2004vq}, 
a~significant enhancement
of the~branching fraction for the~\mbox{\BcToXpi} mode is expected, 
which is anticipated to be similar 
to the~\decay{\Bc}{\jpsi\pip} channel~\cite{PhysRevD.94.034036}. 

In 2017, the ATLAS collaboration measured  the production cross-sections  of the~\chiconex state
in \proton\proton~collisions at~\mbox{$\sqs=8\tev$}~\cite{ATLAS:2016kwu}, where  the short-lived contribution of the nonprompt production was assumed to arise from the decays of~\Bc~mesons.
Subsequently, they reported the fraction of 
the \chiconex,  
produced 
from \Bc~decays, 
integrated over 
the~transverse momentum and 
rapidity range \mbox{$(\pt>10\gevc,\left|y\right|<0.75)$}, to be 

\begin{equation*}
    \dfrac{\upsigma_{\decay{\proton\proton}{\Bc\PX}}
    \times 
    \BR_{ \decay{\Bc}{\chiconex\PX}} } 
   {\upsigma_{\decay{\proton\proton}{\bquark\bquarkbar \PX}}
   \times 
   \BR_{\decay{\bquark}{\chiconex\PX}} } 
   = \left( 25 \pm 13 \pm 2 \pm 5 \right) \%\,,
\end{equation*}
where the~first uncertainty is statistical, the second one systematic 
and the~last  one 
due to the~unknown  polarisation of the~\chiconex~state in $\Bc$\nobreakdash-meson decays. 
Since the~production cross\nobreakdash-section of $\Bc$\nobreakdash-mesons  
is much smaller than the~inclusive beauty production 
cross\nobreakdash-section~\cite{LHCb-PAPER-2014-050},
this result can be interpreted as a~significant enhancement 
of~\chiconex~production in \Bc~meson decays.

This paper reports a~search 
for the~\mbox{$\decay{\Bc}{\chiconex\pip}$}~decay\footnote{Inclusion of charge-conjugate states is implied throughout the paper.}
using $\proton\proton$ collision data, 
corresponding to an~integrated 
luminosity of~9\,\invfb,
collected with the~\lhcb detector 
at~centre-of-mass energies of 7, 8, and 13\,\tev. 
The~decay mode \mbox{$\decay{\Bc}{\psitwos\pip}$}
is used as a~normalisation channel
and the~result reported as a~ratio of 
branching fractions,
\begin{equation}
\mathcal{R}^{\chiconex}_{\psitwos} 
\equiv 
\dfrac{\BR_{\BcToXpi}}{\BR_{\BcToPpi}}
\times\dfrac{\BR_{\ChiconexTojpsipipi}}{\BR_{\decay{\psitwos}{\jpsi\pip\pim}}}\,.  \label{eq:R}
\end{equation}
A~high\nobreakdash-yield 
sample 
of 
\mbox{$\decay{\Bc}{\jpsi\pip\pip\pim}$}
decays~\cite{LHCb-PAPER-2011-044,
LHCb-PAPER-2021-034,
LHCb-PAPER-2022-025},  
without intermediate 
either 
$\chicone(3872)$
or \psitwos~states,
is used as a~control mode 
to calibrate detector resolution effects.
Samples of~\mbox{$\decay{\Bu}{\chiconex\Kp}$}
and \mbox{$\decay{\Bu}{\psitwos\Kp}$}~decays 
are used to parameterise  
the~\mbox{$\decay{\chiconex}{\jpsi\pip\pim}$} and 
 \mbox{$\decay{\psitwos}{\jpsi\pip\pim}$}~signals 
 and 
to study systematic uncertainties.

\section{Detector and simulation}
\label{sec:Detector}

The \lhcb detector~\cite{Alves:2008zz,LHCb-DP-2014-002} is a single-arm forward
spectrometer covering the~pse\-udora\-pi\-dity range \mbox{$2<\eta <5$},
designed for the study of particles containing $\bquark$~or~$\cquark$~quarks. 
The~detector includes a high-precision tracking system consisting of a 
silicon-strip vertex detector surrounding the \proton\proton interaction
region~\cite{LHCb-DP-2014-001}, a large-area silicon-strip detector located
upstream of a dipole magnet with a bending power of about $4{\mathrm{\,T\,m}}$,
and three stations of silicon-strip detectors and straw
drift tubes~\cite{LHCb-DP-2013-003,LHCb-DP-2017-001} placed downstream of the magnet. 
The tracking system provides a measurement of the momentum of charged particles
with a relative uncertainty that varies from $0.5\%$ at low momentum to $1.0\%$~at~$200 \gevc$. 
The~momentum scale is calibrated using samples of $\decay{\jpsi}{\mumu}$ 
and $\decay{\Bu}{\jpsi\Kp}$~decays collected concurrently
with the~data sample used for this analysis~\cite{LHCb-PAPER-2012-048,LHCb-PAPER-2013-011}. 
The~relative accuracy of this
procedure is estimated to be $3 \times 10^{-4}$ using samples of other
fully reconstructed $\bquark$~hadrons, 
$\PUpsilon$~and
$\KS$~mesons.
The~minimum distance between a~track and 
a~primary 
$\proton\proton$\nobreakdash-collision 
vertex\,(PV)~\cite{Bowen:2014tca,Dziurda:2115353}, 
the~impact parameter, 
is~measured with a~resolution of $(15+29/\pt)\mum$, where \pt is 
the~component 
of the~momentum transverse 
to the~beam, in\,\gevc. Different types of charged hadrons
are distinguished using information from 
two ring\nobreakdash-imaging Cherenkov 
detectors\,(RICH)~\cite{LHCb-DP-2012-003}. Photons,
electrons and hadrons are identified 
by a~calorimeter system consisting of scintillating\nobreakdash-pad 
and preshower detectors, 
an electromagnetic and 
a~hadronic calorimeter. Muons are~identified by a~system 
composed of alternating layers of iron and multiwire proportional chambers~\cite{LHCb-DP-2012-002}.

The online event selection is performed by a trigger~\cite{LHCb-DP-2012-004}, 
which consists of a hardware stage, based on information from the calorimeter and muon systems,
followed by a~software stage, which performs a full event reconstruction. 
The~hardware trigger selects muon candidates 
with high transverse momentum 
or dimuon candidates with a~high value of 
the~product
of the~transverse momenta of the~two muons.
In~the~software trigger, 
two 
oppositely charged muons are required to form 
a~common
vertex that is significantly 
displaced from any~PV,
and the~mass of the~$\mumu$~pair 
is required to  
exceed~$2.7\gevcc$.

Simulated events  are used 
to model the~signal mass shapes, 
optimise 
the~selection requirements 
and compute the~efficiencies 
required for determination of 
the~branching fraction ratio
$\mathcal{R}^{\chiconex}_{\psitwos}$.
The~\pythia~\cite{Sjostrand:2007gs} 
generator with a~specific 
\lhcb configuration~\cite{LHCb-PROC-2010-056}
is used to simulate
\proton\proton collisions with~\Bu~meson production.
For~\Bc~meson production,
the~{\sc{BcVegPy}}~generator~\cite{Chang:2003cq,
Chang:2005hq,
Wang:2012ah,
Wu:2013pya}
is used. 
It~is~based on the~full perturbative-QCD
calculations at~lowest 
order $\upalpha_{\mathrm{s}}^4$
via the~dominant 
gluon\nobreakdash-gluon fusion 
processes
\mbox{$\decay{\mathrm{gg}}
{\Bc\left(\B^{*+}_{\cquark}\right)
+ \cquarkbar + \bquark}$}
and 
neglecting contributions from the~quark-pair annihilation 
channels \mbox{$\decay{\mathrm{\quark\quarkbar}}
{\Bc\left(\B^{*+}_{\cquark}\right) + 
\cquarkbar + \bquark}$}~\cite{Chang:1992jb,
Chang:1994aw,
Chang:1996jt,
Kolodziej:1995nv,
Berezhnoy:1996ks}.
The~generator is
interfaced with the~\pythia~parton shower 
and hadronisation model.
\mbox{Decays} of unstable particles are described by            
the~\evtgen 
package~\cite{Lange:2001uf}, 
in which final-state radiation is generated using \photos~\cite{davidson2015photos}. 
The~\mbox{$\decay{\chiconex}{\jpsi\pip\pim}$}~decay 
is modelled as an~$\PS$\nobreakdash-wave decay
via the~intermediate $\jpsi\rhoz$~state~\cite{LHCb-PAPER-2013-001, LHCb-PAPER-2015-015},
while 
the~\mbox{$\decay{\psitwos}{\jpsi\pip\pim}$}~decay 
follows the~model described in 
 Refs.~\cite{Brown:1975dz,Voloshin:2006ce,Voloshin:2007dx}.
Both decay models 
are further corrected using 
large samples of~\mbox{$\decay{\Bu}{\chiconex}\Kp$}
and \mbox{$\decay{\Bu}{\psitwos}\Kp$}~decays~\cite{LHCb-PAPER-2020-009,LHCb-PAPER-2021-045}. 
The~interaction of the~generated particles with the~detector,
and its response, are implemented using
the~\geant toolkit~\cite{Allison:2006ve, *Agostinelli:2002hh} 
as described in Ref.~\cite{LHCb-PROC-2011-006}.
The~transverse momentum and rapidity spectra of 
the~\Bc mesons in simulated samples
are adjusted to match those observed in 
a~high\nobreakdash-yield, 
low\nobreakdash-background sample of reconstructed
\BcTojpsipi decays. 
The~detector response used for the~identification of pions
is sampled from 
the~\mbox{$\decay{\Dstarp}{\left(\decay{\Dz}{\Km\pip}\right)\pip}$}
and \mbox{$\decay{\KS}{\pip\pim}$} control channels~\cite{LHCb-DP-2012-003,
LHCb-DP-2018-001}. To~account for imperfections in the~simulation of
charged-particle reconstruction, 
the~track  reconstruction efficiency
determined from simulation 
is corrected using 
\mbox{$\decay{\jpsi}{\mumu}$}~calibration samples~\cite{LHCb-DP-2013-002}.

\section{Event selection}
\label{sec:Selection} 
The 
\mbox{$\decay{\Bc}{\chiconex\pip}$} and 
\mbox{$\decay{\Bc}{\psitwos\pip}$} candidates
are reconstructed using 
the~\mbox{$\decay{\PX_\ccbar}{\jpsi\pip\pim}$}~decay 
mode.\footnote{The~symbol $\PX_{\cquark\cquarkbar}$ denotes
the~\chiconex and \psitwos~states collectively.}
As~\mbox{$\decay{\Bc}{\PX_{\ccbar}\,\pip}$} and 
\mbox{$\decay{\Bc}{\jpsi\pip\pip\pim}$}~decays 
share the~same final state, 
the~same preselection is applied to all 
three modes. 
For all cases, the~\jpsi~candidates are reconstructed 
via their dimuon decays.

Muon and pion candidates are identified 
by combining information from
the~RICH, calorimeter and muon systems, 
and are required to have transverse momenta
larger than 550\mevc and 
200\mevc for muon and pion candidates, respectively.
Pions are required to have a momentum 
between 3.2 and 150\gevc to ensure good performance 
of particle identification in 
the~RICH detectors~\cite{LHCb-PROC-2011-008,
LHCb-DP-2012-003}. 
To~reduce combinatorial
backgrounds from particles 
produced in \proton\proton~interactions, 
only tracks that are 
inconsistent
with originating from any PV are used.

Pairs of oppositely charged 
muons consistent with originating from a~common vertex 
are combined to form ~\mbox{$\decay{\jpsi}{\mumu}$}~candidates. 
The~reconstructed mass of the~\mumu~pair 
is required to be
in the~range $3.0<m_{\mumu}<3.2\gevcc$. 
The~position of
the~reconstructed dimuon vertex is required to be 
separated from any reconstructed PV.

The~selected $\jpsi$~candidates 
are combined with a~pair of oppositely charged pions 
to form $\PX_{\ccbar}$~candidates.
The~$\jpsi\pip\pim$
mass comprising \psitwos candidates is required 
to be in 
the~\mbox{$3.66<m_{\jpsi\pip\pim}<3.71\gevcc$} region,
while for the~\chiconex~candidates 
the~$3.85<m_{\jpsi\pip\pim}<3.90\gevcc$~region is used. 
The~selected $\PX_\ccbar$ candidates are combined with charged
tracks identified as pions
to form~\mbox{\BcToXccpi}~candidates.
To~construct 
the~\mbox{\BcTojpsitripi}~control sample, 
the~selected $\jpsi$~candidates 
are combined with three charged tracks identified as pions.
The~$\jpsi\pip\pip\pim$~combinations already selected as~\mbox{\BcToXccpi}~candidates 
are not included as 
\mbox{\BcTojpsitripi}~candidates. 

A kinematic fit~\cite{Hulsbergen:2005pu}
that constrains 
the~five\nobreakdash-track 
combination 
to form a~common vertex is performed, in which 
the~mass of the~\mumu~combination is
set equal to the~known \jpsi mass~\cite{PDG2024},  
and the~\Bc~candidate is constrained to originate from 
the~associated~PV.
Each \Bc~candidate 
is associated with the~PV that 
yields the smallest $\chi^2_{\rm{IP}}$, where $\chi^2_{\rm{IP}}$ is defined 
as the~difference in the~vertex\nobreakdash-fit 
$\chi^2$ of a~given PV reconstructed with and
without the~particle under consideration.
A~good fit quality is required to further suppress
combinatorial background. 
The~measured decay time of the~selected candidate
is required to be greater 
than~$75\mum/c$,   
to~reduce the~background from 
particles originating directly from the~PV.

To further reduce 
the combinatorial background
for the~\BcToXccpi and \mbox{\BcTojpsitripi}~decays,
a~multivariate classifier
is used
based on a decision
tree with gradient boosting\,({\sc{BDTG}})~\cite{Breiman}.
The classifier 
is trained on a~mixture of simulated samples from \BcToXpi and \BcToPpi decays
to represent the~signal.
As~a~proxy for the~background, \Bc~candidates from data with 
a~mass between 6.35 and 6.60\gevcc are used, 
excluding the~regions of $m_{\jpsi\pip\pim}$ 
populated by  $\mbox{\decay{\chiconex}{\jpsi\pip\pim}}$
and \decay{\psitwos}{\jpsi\pip\pim}~decays.
The~$k$\nobreakdash-fold
cross\nobreakdash-validation 
technique~\cite{geisser1993predictive} with $k = 13$ 
is taken to avoid 
introducing a~bias in the~{\sc{BDTG}}~evaluation.  
The~classifier is trained
using variables related to the~reconstruction quality, 
decay time of \Bc~candidates,
kinematics of particles in the~final state 
and variables related to pion
identification~\cite{LHCb-PROC-2011-008,
LHCb-DP-2012-003}.
The~requirement on the~{\sc{BDTG}}~output is 
 chosen to maximise 
 the~Punzi figure\nobreakdash-of\nobreakdash-merit
 \mbox{$\upvarepsilon/(\upalpha/2+
 \sqrt{B})$}~\cite{Punzi:2003bu}, 
 where 
 $\upvarepsilon$ is the~signal efficiency 
in simulation for \BcToXpi decay,
$\upalpha = 5$ 
 represents the~desired signal significance
 in~units of standard deviations, and  
 $B$~is the~expected background     
 yield  within a~narrow mass region
 centred at the~known $\Bc$~and~$\chiconex$
 masses~\cite{PDG2024}.

\begin{figure}[t]
	\setlength{\unitlength}{1mm}
	\centering
	\begin{picture}(160,185)
	\put( 0,0){\includegraphics*[width=160mm]{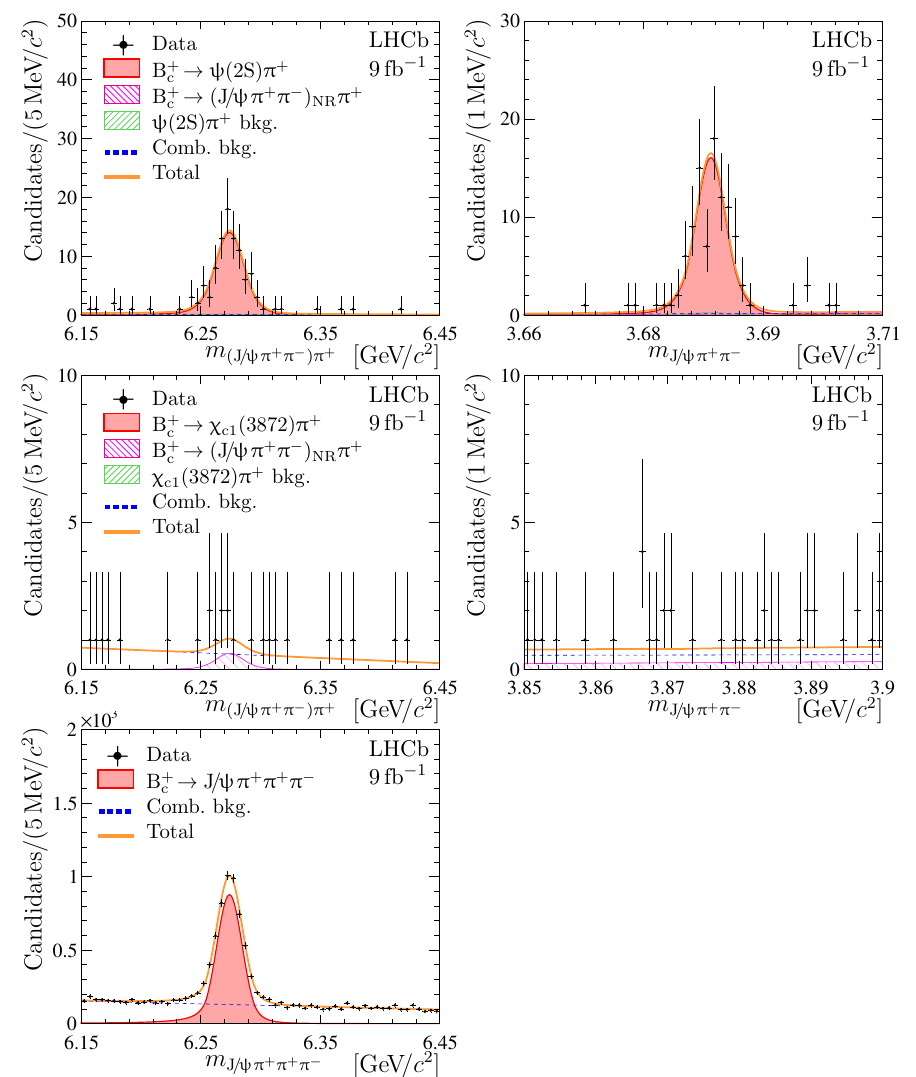}}
	\end{picture}
	\caption {\small 
	Distributions 
       and fit projections of 
	the (left) \jpsi\pip\pip\pim and (right)
	\JpsiPiPi masses
	for selected 
        (top) \mbox{\BcToPpi}, 
	(middle) \mbox{\BcToXpi}
        and (bottom) \mbox{\BcTojpsitripi} candidates. 
For \mbox{$\decay{\Bc}{\PX_{\ccbar}\pip}$}~decays,       
the ($\jpsi\pip\pim)\pip$  and $\jpsi\pip\pim$  mass 
spectra are shown for candidates within narrow regions 
around the known $\PX_\ccbar$ and $\Bc$~masses, 
respectively, as described in the~text.}
	\label{fig:signal_fit_Bc}
\end{figure}

The~mass distributions for selected 
\BcToXccpi and \mbox{\BcTojpsitripi} candidates 
are shown in~Fig.~\ref{fig:signal_fit_Bc}.
To~improve the~resolution on 
the~$\jpsi\pip\pim$~mass
for the~\BcToXccpi~candidates 
and to eliminate a~small correlation between 
$m_{\left(\jpsi\pip\pim\right)\pip}$ and $m_{\JpsiPiPi}$~~\cite{LHCb-PAPER-2020-009,LHCb-PAPER-2020-035},
the~$m_{\JpsiPiPi}$ variable is computed
by constraining the~mass of the~\Bc~candidates
to its~known mass~\cite{PDG2024}. 
For~better visibility, 
the~$\left(\jpsi\pip\pim\right)\pip$~mass spectra 
for the~\mbox{$\decay{\Bc}{\PX_{\ccbar}\,\pip}$}~candidates
are shown for candidates within 
the~narrow mass regions
around the~known masses 
of the~$\psitwos$ and 
$\chiconex$~mesons,
\mbox{$3.680<m_{\jpsi\pip\pim}<3.692\gevcc$} 
and 
\mbox{$3.864<m_{\jpsi\pip\pim}<3.879\gevcc$}, respectively.  
The~$\jpsi\pip\pim$~mass spectra
are shown for candidates within
the~narrow mass region around 
the~known mass of 
the~\Bc~meson
\mbox{$6.24<m_{\left(\jpsi\pip\pim\right)\pip}<6.31\gevcc$}.

\section{Signal yields and efficiencies}
\label{sec:Sig_eff}

The yields for the~\BcToXccpi and \BcTojpsitripi~decays 
are determined 
using a~simultaneous
extended unbinned maximum\nobreakdash-likelihood fit
to the 
two\nobreakdash-dimensional distributions
of $m_{\left(\jpsi\pip\pim\right)\pip}$ \vs 
$m_{\jpsi\pip\pim}$
for
the~\BcToXccpi~candidates, 
and the one-dimensional mass distribution of selected 
\mbox{\BcTojpsitripi}~control candidates. 


The two-dimensional fit functions for 
the~\mbox{\BcToXccpi} channels 
are defined as a~sum of four components: 
\begin{enumerate}
	\item Signal  \decay{\Bc}{(\decay{\PX_\ccbar}{\JpsiPiPi})\pip} decays
	parameterised as a~product of  \Bc~and $\PX_\ccbar$~signal shapes, 
        each of which are modelled by the~sum of a~Gaussian function with 
        a~modified Gaussian function 
        with power\nobreakdash-law\nobreakdash-replaced 
        tails on both sides of the~distribution~\cite{LHCb-PAPER-2011-013,Skwarnicki:1986xj}; 
	
        \item 
        Contributions 
	from nonresonant decays \mbox{\decay{\Bc}{\left(\jpsi\pip\pim\right)_{\mathrm{NR}}\pip}}
	without proceeding through 
	a~narrow intermediate $\PX_\ccbar$~state,
	parameterised as a~product of the~\Bc~signal shape with 
        a three-body phase-space function~\cite{Byckling}
        and a linear function in 
        the~$\jpsi\pip\pim$~mass;

	\item Random combinations of  
	$\PX_\ccbar$ and \pip~candidates,
         parameterised as a~product of the~$\PX_\ccbar$~signal shape 
        with a~linear function in the~$\left(\jpsi\pip\pim\right)\pip$~mass;
        
	\item 
    Combinatorial background 
    consisting of random $\left(\jpsi\pip\pim\right)\pip$ combinations, 
	described by a~two\nobreakdash-dimensional 
        nonfactorisable 
	positive 
        second\nobreakdash-order polynomial function.
\end{enumerate}
For~the \BcTojpsitripi~channel, the~one-dimensional 
fit function consists of two components: 
\begin{enumerate}
	\item Signal \BcTojpsitripi decays,
	parameterised by the~modified Gaussian function previously described; 
    \item 
    Combinatorial background, 
    modelled by a~first\nobreakdash-order polynomial function.
\end{enumerate}
The~shape parameters for the~fit components describing~\mbox{\BcToXccpi} and \mbox{\BcTojpsitripi}~decays 
are taken from simulation and their uncertainties 
propagated to the~fit through multivariate Gaussian constraints.  
The~\Bc~signal peak position parameter 
is shared between all decays and allowed to vary in the~fit. 
The~mass resolution parameters $\sigma_{\Bc}$ 
for 
the~\Bc~components 
are parameterised as 
\begin{equation}
   \sigma_{\Bc}  = f_{\Bc} \times  
   \sigma_{\Bc}^{\mathrm{sim}} \,, 
\end{equation}
where
$\sigma_{\Bc}^{\mathrm{sim}}$~are~mass resolution parameters
obtained from simulation.
The~parameter 
$f_{\Bc}$
accounts for a~possible 
difference between data and 
simulation. 
Furthermore, the~parameter  $f_{\Bc}$ is shared and 
free to vary in the~fit to data.
The~position and mass resolution parameters for 
the~\mbox{$\PX_\ccbar$}~components,
$m_{\chiconex}$,
$m_{\psitwos}$, 
$\sigma_{\chiconex}$, and 
$\sigma_{\psitwos}$,  
are parameterised~as 
\begin{subequations} \label{eq:params_Bu}
\begin{eqnarray}
  m_{\chiconex}     & = &  \updelta m_{\PX_{\ccbar}} + m_{\psitwos} \,,  
   \\ 
   \sigma_{\chiconex} & = &  f_{\PX_\ccbar} \times  \sigma_{\chiconex}^{\mathrm{sim}} \,, 
   \\
   \sigma_{\psitwos} & = &  f_{\PX_\ccbar} \times  \sigma_{\psitwos}^{\mathrm{sim}} \,, 
\end{eqnarray} \label{eq:Bu_params}\end{subequations}
where the~position~of the~\psitwos~signal peak, $m_{\psitwos}$, 
is allowed to vary in the fit, while 
$\sigma_{\chiconex}^{\mathrm{sim}}$ 
and $\sigma_{\psitwos}^{\mathrm{sim}}$~are 
mass resolution parameters
obtained from simulation.
The~parameter 
$\updelta m_{\PX_{\ccbar}}$ 
corresponds to the mass difference
between the~\chiconex and \psitwos mesons, 
while
$f_{\PX_\ccbar}$
describes potential
differences between data and 
simulation in terms of~resolution. 
Both 
parameters are determined in Ref.~\cite{LHCb-PAPER-2020-009} 
and their uncertainties propagated to the fit 
via Gaussian constraints.

The~yield of 
the~\BcToXpi~signal component,  
$N_{\decay{\Bc}{\chiconex\pip}}$, 
is parameterised~as
\begin{equation}
\label{eq:yield_chic1}
    N_{\decay{\Bc}{\chiconex\pip}} \equiv 
    N_{\decay{\Bc}{\psitwos\pip}} 
    \times 
    \mathcal{R}^{\chiconex}_{\psitwos}
    \times
    \upvarepsilon^{\chiconex}_{\psitwos}   \,, 
\end{equation}
where
$N_{\decay{\Bc}{\psitwos\pip}}$
is
the~yield
of the~\BcToPpi~signal 
component,   
$\mathcal{R}^{\chiconex}_{\psitwos}$ is 
a~free parameter in the~fit corresponding to 
the~ratio of branching fractions
for~\mbox{$\decay{\Bc}{\chiconex\pip}$}
and \mbox{$\decay{\Bc}{\psitwos\pip}$}~decays as defined in Eq.~\eqref{eq:R},  
and $\upvarepsilon^{\chiconex}_{\psitwos}$
is the~ratio of the~total efficiencies
for~\mbox{\BcToXpi} and \mbox{\BcToPpi}~decays,
\begin{equation}
    \upvarepsilon^{\chiconex}_{\psitwos} \equiv 
    \dfrac
    { \upvarepsilon_{\decay{\Bc}{\chiconex\pip}}}
    { \upvarepsilon_{\decay{\Bc}{\psitwos\pip}}}\,.  \label{eq:e12}
\end{equation}
For each channel, the~total efficiency is defined 
as the~product of the~detector acceptance, reconstruction,
selection and trigger efficiencies, 
where each subsequent efficiency 
is defined with
respect to the~previous one. 
Each~of the~efficiencies is calculated using 
the~calibrated simulation samples
described in Sec.~\ref{sec:Detector}.
The~ratio of the~efficiencies 
for the~\mbox{$\BcToXpi$} and 
\mbox{$\BcToPpi$}~channels,
$\upvarepsilon^{\chiconex}_{\psitwos}$,
defined in Eq.~\eqref{eq:e12}, 
is found to be 
\begin{equation*}
 \upvarepsilon^{\chiconex}_{\psitwos} =  
 \ 1.673 \pm 0.003 \,, \label{eq:effs}
 \end{equation*}
where the~uncertainty is due to size of 
the~simulated samples. 
A~smaller energy release in 
the~\mbox{\decay{\psitwos}{\jpsi\pip\pim}}~decay
causes a~larger fraction of charged pions 
to escape reconstruction and selection, 
that is a~primary reason for 
the~smaller  efficiency for 
the~\mbox{$\decay{\Bc}{ \psitwos\pip}$}
channel~\cite{LHCb-PAPER-2019-023,
LHCb-PAPER-2020-009,
LHCb-PAPER-2020-035,
LHCb-PAPER-2022-049}.
The~uncertainties for
the~$\upvarepsilon^{\chiconex}_{\psitwos}$~value, 
including~systematic effects, 
discussed in detail in Sec.~\ref{sec:Systematics}, 
is  
accounted for in the~fit via a~Gaussian constraint. 

The~result of the~fit is shown
in Fig.~\ref{fig:signal_fit_Bc}, and 
the~parameters of interest 
summarised in Table~\ref{tab:sim_fit_res}.
The~\Bc~peak position parameter 
$m_{\Bc}$ is in 
good agreement with 
the~known mass of 
the~\Bc~meson~\cite{PDG2024},
and the~resolution scale factor
$f_{\Bc}$ is in good agreement
with the~value obtained in Ref.~\cite{LHCb-PAPER-2021-034}.
The~\BcToXpi~signal 
is observed 
to be consistent with zero. 


\begin{table}[t]
	\centering
	\caption{\small 
	Parameters of interest obtained from the~simultaneous  
	unbinned extended 
	maximum\protect\nobreakdash-likelihood fit.
    In~addition to statistical sources, 
    the~uncertainties also account for 
    a~systematic component due to 
    the~inclusion of $\upvarepsilon^{\chiconex}_{\psitwos}$ in the fit.
	}
	\begin{tabular*}{0.5\textwidth}
	{@{\hspace{5mm}}l@{\extracolsep{\fill}}cc@{\hspace{5mm}}}
    \multicolumn{2}{l}{Parameter} &  Value \\[1.5mm]   
       \hline     \\[-3mm]
        \multicolumn{2}{l}{\ \ $N_{\decay{\Bc}{\jpsi\pip\pip\pim}}$}   &  \ $5049\pm91$  \\
        \multicolumn{2}{l}{\ \ $N_{\decay{\Bc}{\psitwos\pip}}$}        &  \ \phantom{0} $96 \pm 11$ 
       \\ 
    $\mathcal{R}^{\chiconex}_{\psitwos}$ 
    & $\left[\%\right]$  
    & $\ -0.1 ^{\,+\,2.5}_{\,-\,2.1}$ 
    \\ 
     $m_{\Bc}$ &  $\left[\!\mevcc\right]$      & $6274.15 \pm 0.21$  \\
     $m_{\psitwos}$  &$\left[\!\mevcc\right]$ & $3686.05 \pm 0.27$ \\
     $\updelta m_{\PX_\ccbar}$ &$\left[\mevcc\right]$ &\, $185.54  \pm 0.06$ \\
      $f_{\Bc}$             &              &\phantom{00,} $1.118\pm0.021$\\
      $f_{\PX_\ccbar}$      &               &\phantom{00,} $1.048\pm0.004$\\

     \end{tabular*}
     \label{tab:sim_fit_res}
\end{table}

\section{Systematic uncertainties}
\label{sec:Systematics}

The decay channels 
under investigation 
share similar kinematics and topologies, 
leading to the cancellation of many sources of systematic uncertainties 
in the branching-fraction ratio. 
The~remaining contributions 
to the~systematic uncertainty are 
discussed below.
There are two categories of systematic uncertainties. 
The~first category consists of  uncertainties 
that affect the~ratio of~total efficiencies
$\upvarepsilon^{\chiconex}_{\psitwos}$ from Eq.~\eqref{eq:e12}, 
mostly related to the difference between data and simulation.
Those are summarised in Table~\ref{tab:systematics}.
The~second category consists of 
uncertainties  
related to the choice of the fit model.


To address the differences between 
data and 
simulation,
the~transverse momentum and rapidity spectra of~\Bc mesons in the~simulated samples 
are adjusted to match those observed 
in a~high\nobreakdash-yield 
low\nobreakdash-background
sample of \BcTojpsipi decays~\cite{LHCb-PAPER-2023-039}. 
The~finite size of this sample introduces  
uncertainty in the~resulting production
spectra of the~\Bc~mesons. 
The~associated systematic uncertainty in the~efficiency
ratio
is estimated using the~variation of 
the~\Bc~kinematic spectra 
within their uncertainties, for which 
the~induced relative variation 
on the~efficiency ratio 
$\upvarepsilon^{\chiconex}_{\psitwos}$
is found to be much smaller than 0.1\%.

Due to the~slightly different kinematic distributions of
the~final\nobreakdash-state particles,
there are residual differences in 
the~reconstruction 
efficiency of charged\nobreakdash-particle tracks
that do not completely cancel out in the~efficiency ratio.
The~track\nobreakdash-finding 
efficiency obtained from simulated samples
is corrected using calibration 
channels~\cite{LHCb-DP-2013-002}. 
Uncertainties related to such~efficiency
correction factors are then propagated to the~ratio 
of total 
efficiencies using pseudoexperiments
and found to be 
0.2\%.

Pion identification 
in the~simulation is modelled by sampling 
the~corresponding distributions 
in the~\mbox{$\decay{\Dstarp}
{\left(\decay{\Dz}{\Km\pip}\right)\pip}$}
and \mbox{$\decay{\KS}{\pip\pim}$}~control 
channels~\cite{LHCb-DP-2012-003,
LHCb-DP-2018-001}. 
The~systematic uncertainty obtained
through this procedure originates from the~kernel 
shape used in the~estimation of the~probability-density distributions. An~alternative
response is estimated using 
a~different 
kernel estimation with a~modified shape, and 
the~efficiency
models recomputed~\cite{LHCb-PAPER-2020-025, Poluektov:2014rxa}.
The~difference between the~two estimates
for the~efficiency ratio is taken as the~systematic
uncertainty related to pion identification 
and is found to be 
0.1\%.

Large samples of~\mbox{$\decay{\Bp}{\jpsi\Kp}$} and 
\mbox{$\decay{\Bp}{\psitwos\Kp}$}~decays~\cite{LHCb-PAPER-2012-010}
are used to estimate the~systematic uncertainty 
related to the~trigger efficiency. 
A~conservative estimate of 
1.1\% for the~relative difference 
between data and simulation is taken
as the~corresponding systematic uncertainty~\cite{LHCb-PAPER-2012-010}.

Discrepancies in reconstructed quantities between 
the~simulated samples and  data,
arising from factors other than those previously described, 
are studied by varying the~{\sc{BDTG}}~selection criteria
for the~control 
\mbox{$\decay{\Bp}{\chiconex\Kp}$}
and \mbox{$\decay{\Bp}{\psitwos\Kp}$} decays~\cite{LHCb-PAPER-2020-009} 
over the~entire range. 
The~observed maximal difference between 
the~efficiency estimated using 
data and simulation does not exceed 
0.8\%, 
which 
is taken as a~corresponding 
systematic uncertainty.

Model corrections for 
$\decay{\chiconex}{\jpsi\pip\pim}$
and $\decay{\psitwos}{\jpsi\pip\pim}$~decays are obtained from the~control 
\mbox{$\decay{\Bp}{\chiconex\Kp}$}
and \mbox{$\decay{\Bp}{\psitwos\Kp}$}
decay modes~\cite{LHCb-PAPER-2020-009} 
using the algorithm described in Ref.~\cite{Rogozhnikov:2016bdp}.
The~systematic uncertainty associated with the correction is
estimated as the largest deviation
of the~efficiency ratio  $\upvarepsilon^{\chiconex}_{\psitwos}$
value from the baseline and found to be 
1.5\%,
which is assigned as
systematic uncertainty
associated with the~$\PX_\ccbar$ decay model.

The~systematic uncertainty 
due to the~finite size of the~simulated samples,
used to calculate the~efficiency ratio
from Eq.~\eqref{eq:e12}, 
is found to be  
0.2\%. 
The~systematic uncertainties
for the~efficiency ratio 
$\upvarepsilon^{\chiconex}_{\psitwos}$
are summarised in Table~\ref{tab:systematics}, where 
their total
is estimated as the~sum in quadrature of 
the~individual contributions
and accounted for in the~fit via a~Gaussian constraint. 

\begin{table}[tb] 
	\centering
	\caption{\small
    Relative systematic uncertainties (in \%) for the efficiency ratio of  the~\mbox{$\decay{\Bc}{\chiconex\pip}$}
    and \mbox{$\decay{\Bc}{\psitwos\pip}$}~decays.
    The~total uncertainty is calculated as the~quadratic sum of individual contributions.} 
	\label{tab:systematics}
	\vspace{2mm}
	\begin{tabular*}{0.70\textwidth}
	{@{\hspace{5mm}}l@{\extracolsep{\fill}}c@{\hspace{5mm}}}
	Source  
        & 
        $\sigma \left( \upvarepsilon^{\chiconex}_{\psitwos}\right)\,\left[\%\right]$
   \\[2.5mm]
  \hline 
  \\[-3.5mm]
  \Bc~production spectra                 &   $\!<0.1\phantom{<}$ 
  \\
  Track reconstruction                   &    $0.2$ 
  \\
  Pion identification                    &    $0.1$ 
  \\ 
  Trigger efficiency                     &     $  1.1$ 
  \\
  Data-simulation difference             &  $0.8$ 
  \\
  $\chiconex$ and  \psitwos~decay models   & $1.5$ 
  \\
  Size of simulated samples              & $ 0.2$ 
    \\[1.5mm]
  \hline 
  \\[-3.5mm]
  Sum in quadrature   &   $2.0$ 
 
	\end{tabular*}
	\vspace{3mm}
\end{table}
Using the~$\mathrm{CL_s}$ technique~\cite{CLs}, 
where the~$p$\nobreakdash-values are 
computed based on the~asymptotic 
properties of the~profile likelihood 
ratio~\cite{Cowan:2010js}, 
an~upper limit of
$\mathcal{R}^{\chiconex}_{\psitwos} < 0.04\,(0.05) \,$
is set at the 90\,(95)\% confidence level\,(CL), 
accounting for the systematic effects relating to the efficiency ratio.
The~remaining uncertainty 
is related to the~imperfect understanding 
of the~shapes of the~signal and background 
components used in the~fits. 
To~estimate this, several alternative signal shapes  
are tested, in particular, 
the~sum of a~generalised Student's
$t$\nobreakdash-distribution~\cite{Jackman} 
with a~Gaussian function, 
and the~sum of  
a~modified 
Apollonios function~\cite{Santos:2013gra} with
a~Gaussian function.
For~the~background components, 
the~product of an~exponential function 
with a~positive first\nobreakdash-order 
polynomial function, 
as well as a~positive 
second\nobreakdash-order polynomial function, 
are tested as alternative background shapes 
for the~fit to the $\PX_\ccbar\pip$ and $\jpsi\pip\pim$~mass spectra.
For each alternative model,
 the~upper limit on
the~$\mathcal{R}^{\chiconex}_{\psitwos}$~ratio 
is recalculated where 
the~maximal obtained value 
\begin{equation*}
   \mathcal{R}^{\chiconex}_{\psitwos}    
    < 0.05\,(0.06)~\text{at 90\,(95)\%\ CL} \,,
\end{equation*}
is conservatively taken as the~upper limit including 
systematic uncertainties. 

\section{Conclusions}
\label{sec:Results}

\begin{figure}[tbh]
	\setlength{\unitlength}{1mm}
	\centering
	\begin{picture}(150,120)
	\put(0,0){\includegraphics*[width=160mm,height=120mm]{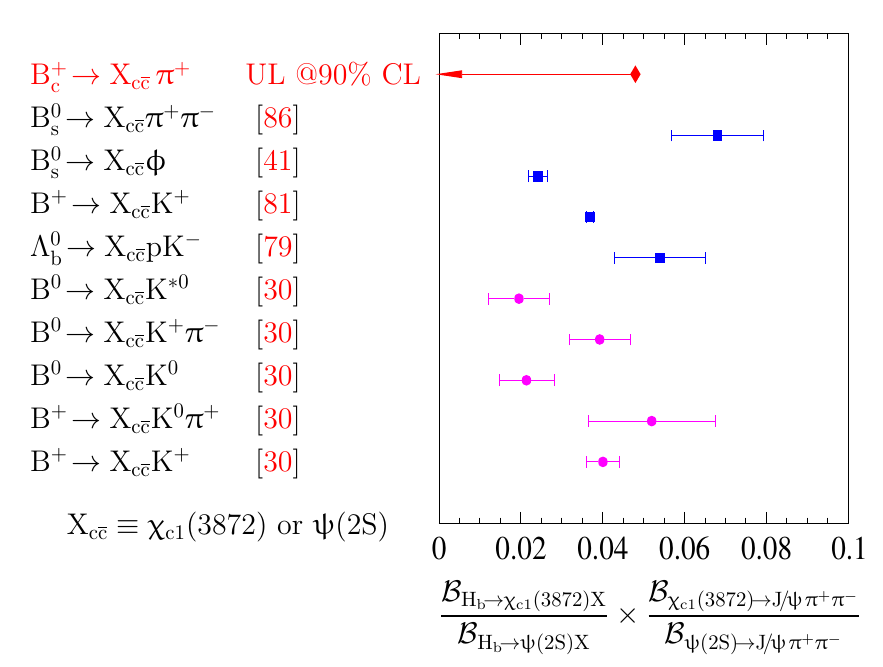}}
	\end{picture}
    \caption{\small  
    Comparison of the upper limit on the branching fraction 
    ratio between \mbox{\BcToXpi} and \BcToPpi 
    decays set in this analysis, shown in red,  
    with measurements 
    of the~ratios of branching fractions
    for~\mbox{$\decay{\PH_\bquark}{\chiconex\PX}$}
    and~\mbox{$\decay{\PH_\bquark}{\psitwos\PX}$}
    decays of various beauty hadrons
    performed by the~LHCb collaboration\,(blue squares) and 
    Belle, BaBar and CDF collaborations\,(magenta circles).}
\label{fig:ul_cmp}
\end{figure}

A~search for the~\mbox{$\decay{\Bc}{\chiconex\pip}$}~decay 
is performed using  $\proton\proton$~collision data, 
corresponding to an~integrated luminosity of 
$9\invfb$,
collected with the~\lhcb detector at 
centre\nobreakdash-of\nobreakdash-mass energies of 7, 8, and 13\tev. 
No~\mbox{$\decay{\Bc}{\chiconex\pip}$}~signal 
is observed.
Using~\mbox{$\decay{\Bc}{\psitwos\pip}$}~decays
as a normalisation channel,  
an~upper limit on the~relative branching fraction between~\BcToXpi and \BcToPpi decays is set to be
\begin{equation*}
 \mathcal{R}^{\chiconex}_{\psitwos}  = 
\dfrac{\BR_{\BcToXpi}}{\BR_{\BcToPpi}}
\times \dfrac{\BR_{\ChiconexTojpsipipi}}
{\BR_{\decay{\psitwos}{\jpsi\pip\pim}}} 
 < 0.05\,(0.06)~\text{at 90\,(95)\%\,CL}\,.
 \end{equation*}
A~comparison of this upper limit with similar measurements in
 other beauty-hadron decays, as reported
 in Refs.~\cite{
    LHCb-PAPER-2022-049,
    LHCb-PAPER-2020-035,
    LHCb-PAPER-2020-009,
    LHCb-PAPER-2019-023,
    PDG2024},
is shown in Fig.~\ref{fig:ul_cmp}. 
No~large enhancement for~\chiconex production in \Bc decays is observed, 
which supports neither 
the~conclusions drawn by 
the~ATLAS collaboration~\cite{ATLAS:2016kwu},
nor expectations from the~compact\nobreakdash-tetraquark interpretation
of the~\chiconex~state~\cite{PhysRevD.94.034036}. 

\section*{Acknowledgements}
%
%
\noindent We express our gratitude to our colleagues in the CERN
accelerator departments for the excellent performance of the LHC. We
thank the technical and administrative staff at the LHCb
institutes.
We acknowledge support from CERN and from the national agencies:
ARC (Australia);
CAPES, CNPq, FAPERJ and FINEP (Brazil); 
MOST and NSFC (China); 
CNRS/IN2P3 (France); 
BMBF, DFG and MPG (Germany); 
INFN (Italy); 
NWO (Netherlands); 
MNiSW and NCN (Poland); 
MCID/IFA (Romania); 
MICIU and AEI (Spain);
SNSF and SER (Switzerland); 
NASU (Ukraine); 
STFC (United Kingdom); 
DOE NP and NSF (USA).
We acknowledge the computing resources that are provided by ARDC (Australia), 
CBPF (Brazil),
CERN, 
IHEP and LZU (China),
IN2P3 (France), 
KIT and DESY (Germany), 
INFN (Italy), 
SURF (Netherlands),
Polish WLCG (Poland),
IFIN-HH (Romania), 
PIC (Spain), CSCS (Switzerland), 
and GridPP (United Kingdom).
We are indebted to the communities behind the multiple open-source
software packages on which we depend.
Individual groups or members have received support from
Key Research Program of Frontier Sciences of CAS, CAS PIFI, CAS CCEPP, 
Fundamental Research Funds for the Central Universities,  and Sci.\ \& Tech.\ Program of Guangzhou (China);
Minciencias (Colombia);
EPLANET, Marie Sk\l{}odowska-Curie Actions, ERC and NextGenerationEU (European Union);
A*MIDEX, ANR, IPhU and Labex P2IO, and R\'{e}gion Auvergne-Rh\^{o}ne-Alpes (France);
Alexander-von-Humboldt Foundation (Germany);
ICSC (Italy); 
Severo Ochoa and Mar\'ia de Maeztu Units of Excellence, GVA, XuntaGal, GENCAT, InTalent-Inditex and Prog.~Atracci\'on Talento CM (Spain);
SRC (Sweden);
the Leverhulme Trust, the Royal Society and UKRI (United Kingdom).

\clearpage 


\usetikzlibrary{patterns}

\clearpage
\addcontentsline{toc}{section}{References}
\bibliographystyle{LHCb}
\bibliography{main,standard,LHCb-PAPER,LHCb-CONF,LHCb-DP,LHCb-TDR}
 
\newpage

\centerline
{\large\bf LHCb collaboration}
\begin
{flushleft}
\small
R.~Aaij$^{38}$\lhcborcid{0000-0003-0533-1952},
A.S.W.~Abdelmotteleb$^{57}$\lhcborcid{0000-0001-7905-0542},
C.~Abellan~Beteta$^{51}$,
F.~Abudin{\'e}n$^{57}$\lhcborcid{0000-0002-6737-3528},
T.~Ackernley$^{61}$\lhcborcid{0000-0002-5951-3498},
A. A. ~Adefisoye$^{69}$\lhcborcid{0000-0003-2448-1550},
B.~Adeva$^{47}$\lhcborcid{0000-0001-9756-3712},
M.~Adinolfi$^{55}$\lhcborcid{0000-0002-1326-1264},
P.~Adlarson$^{83}$\lhcborcid{0000-0001-6280-3851},
C.~Agapopoulou$^{14}$\lhcborcid{0000-0002-2368-0147}
% ...

{\footnotesize \it

% ...
$^{14}$Universit{\'e} Paris-Saclay, CNRS/IN2P3, IJCLab, Orsay, France\\
% ...
$^{38}$Nikhef National Institute for Subatomic Physics, Amsterdam, Netherlands\\
% ...
$^{47}$Instituto Galego de F{\'\i}sica de Altas Enerx{\'\i}as (IGFAE), Universidade de Santiago de Compostela, Santiago de Compostela, Spain\\
% ...
$^{51}$Physik-Institut, Universit{\"a}t Z{\"u}rich, Z{\"u}rich, Switzerland\\
% ...
$^{55}$H.H. Wills Physics Laboratory, University of Bristol, Bristol, United Kingdom\\
% ...
$^{57}$Department of Physics, University of Warwick, Coventry, United Kingdom\\
% ...
$^{60}$School of Physics and Astronomy, University of Glasgow, Glasgow, United Kingdom\\
$^{61}$Oliver Lodge Laboratory, University of Liverpool, Liverpool, United Kingdom\\
% ...
$^{69}$Syracuse University, Syracuse, NY, United States\\
% ...
$^{83}$Department of Physics and Astronomy, Uppsala University, Uppsala, Sweden, associated to $^{60}$\\
% ...
}
\end{flushleft}

\end{document}